\documentstyle[preprint,aps]{revtex}

\tightenlines

\begin{document}

\draft

\title{\bf Multi-layer $S={1\over 2}$ Heisenberg antiferromagnet}
\author{Zheng Weihong\cite{byline1}
} 
\address{School of Physics,                                              
The University of New South Wales,                                   
Sydney, NSW 2052, Australia.}                      


\maketitle 

\begin{abstract}
The multi-layer $S={1\over 2}$ square lattice Heisenberg antiferromagnet 
with up to 6 layers is studied via various series expansions. 
For the systems with an odd number of coupled planes, 
the ground-state energy, staggered magnetization, 
and triplet excitation spectra are calculated via two 
different Ising expansions.
The systems are found to have long range N\'eel order and  
gapless excitations for all ratios of interlayer to intralayer 
couplings, as for the single-layer system.
For the systems with  an even number of coupled planes,
there is a second order transition point
separating the gapless Ne\'el phase and gapped quantum disordered spin
liquid phase, and the critical points are located via expansions
in the interlayer exchange coupling. This transition point 
is found to vary about inversely as the number of layers.
The triplet excitation
spectra are also computed, and at the critical point
the normalized spectra
appear to follow a universal function,
independent of number of layers.
\end{abstract}                                                              


\narrowtext
\section{INTRODUCTION}
Low-dimensional quantum antiferromagnets exhibit many
remarkable properties, and the study of these systems
has been  the subject
of intense theoretical and experimental research in recent years.
It is by now well established that one-dimensional Heisenberg antiferromagnets
with integer spin have a gap in the excitation spectrum, whereas 
those with half-integer spin  have gapless excitations.
The former have a finite correlation length, while
for the latter it is infinite with the spin-spin
correlation function decaying to zero as a power law.
In 2-dimensions, the unfrustrated square-lattice Heisenberg model
has long range N\'{e}el order in the ground state. It has gapless
Goldstone modes as expected.
For the 3-dimensional Heisenberg model, one expects a stronger
long-range Ne\'el order\cite{zwh3d} in the ground state.

In recent years much interest has focussed on systems
with intermediate dimensionality and on questions of crossovers
between $d=1$ and $d=2$. The investigations\cite{ladder} showed
that the crossover from the single Heisenberg chain to the
two-dimensional antiferromagnetic  square lattice, obtained
by assembling chains to form a ``ladder'' of increasing width,
is far from smooth. Heisenberg ladders with an even
number of legs (chains) show a completely different behavior
from odd-leg-ladders. While even-leg-ladders have a spin gap
and short range correlations, odd-leg-ladders have no
gap and power-law correlations.

What happens to the crossover
from the two-dimensional antiferromagnetic 
square lattice to the
three-dimensional antiferromagnetic simple cubic lattice, obtained
by assembling planes to form a multi-layer system?
The spin-$S$ bilayer Heisenberg antiferromagnet  has been well studied,
and it has a second order transition\cite{zwh_s,zwh_bilayer}
separating the Ne\'el phase and dimer phase, and this
critical points can be remarkably
well fitted by the following linear relation:
\begin{equation}
(J_2/J_1)_c \simeq 3.72 [ S(S+1) - 0.068]~.
\end{equation}
The systems with more than two planes have not been studied,
as far as we are aware.
As in the case with even and odd numbers of coupled chains,
we may expect the Heisenberg antiferromagnet on 
an odd number of coupled
planes to show  a fundamentally different behavior
from that for an even number of coupled planes.
The multilayer Heisenberg antiferromagnet has recently attracted
attention also due to  its relevance
to the understanding  of the magnetic properties of 
cuprate superconductors (such as YBa$_2$Cu$_3$O$_{6+x}$)
containing two or more  weakly coupled CuO$_2$ layers\cite{rez,hay,mil96}. 

In this paper, we study the zero temperature properties of the
$n$-layer, $S={1\over 2}$, square lattice Heisenberg antiferromagnet, where
each layer is composed of a nearest-neighbor Heisenberg model on a square
lattice, and there is a further antiferromagnetic coupling 
between corresponding sites of each layer.
The system can be described
by the following Hamiltonian:
\begin{equation}
H = J_1 H_{\rm plane} + J_2 H_{\rm rung} \label{H}
\end{equation}
where
\begin{eqnarray}
H_{\rm plane} &=& \sum_{\alpha=1}^n 
\sum_{\langle i,j\rangle} 
{\bf S}_{\alpha,i} \cdot {\bf S}_{\alpha,j}
   \nonumber \\ && \\
H_{\rm rung} &=& \sum_i \sum_{\alpha=1}^{n-1} {\bf S}_{\alpha,i} \cdot {\bf S}_{\alpha+1,i} \nonumber
\end{eqnarray}
${\bf S}_{\alpha,i}$ is a S=1/2 spin operator at 
site $i$ of the $\alpha$th layer, and 
$\langle i,j\rangle$ denotes a pair of nearest neighbor sites on
a square lattice. Here $J_1$ is the interaction between 
nearest neighbor spins in one plane, 
and $J_2$ is the interaction between
adjacent spins from different layers.
We denote the ratio of couplings as $y$,
that is, $y=J_2/J_1$.
In the present paper, we study only the case of antiferromagnetic coupling, 
that is, both $J_1$ and $J_2$ are positive.
In the small $J_2/J_1$ limit,
the model describes $n$ weakly interacting 2D Heisenberg antiferromagnets,
where each of them is  N\'{e}el ordered and possesses gapless Goldstone
excitations. 
While in the large $J_2/J_1$ limit, the system reduces to weakly coupled rungs,
and each rung is described by the Hamiltonian
\begin{equation}
h_{\rm rung} = \sum_{\alpha=1}^{n-1} {\bf S}_{\alpha} \cdot {\bf S}_{\alpha+1}~.
\end{equation}
The ground state of $h_{\rm rung}$ is an $S=0$ singlet for even
$n$, or an $S=1/2$ doublet for odd  $n$. This implies
that there are some fundamental differences between even and odd numbers
of coupled planes.
For an odd number of layers, the system can be mapped  to the well-known
 single layer Heisenberg
antiferromagnet at the large $J_2$ limit, and so the system should have 
Ne\'el order for all ratios of interlayer to intralayer  couplings.
For an even number of layers, on the other hand,
the system has a definite gap at the
large $J_2$ limit, so we expect there is a transition
separating the gapless Ne\'el phase and  gapped 
quantum disordered spin liquid
phase at a certain critical ratio of $J_2/J_1$, 
and this transition should lie in
the universality class of the classical $d=3$ Heisenberg model, as
for the bilayer system\cite{zwh_bilayer}.

To confirm the above arguments,  we study these 
systems with up to 6 layers via various series
expansions.
This paper is organized as follows.
In Section II, we develop two different Ising expansions  
for the 3 and 5-layer systems.
The first one is the usual naive Ising expansion
where we take the $z$-component of the 
Hamiltonian as the unperturbed
Hamiltonian, and the rest of it as the perturbation.
In the second Ising expansion, we choose
the eigenstates of $h_{\rm rung}$, which has an $S=1/2$ 
doublet as the ground state,
as the basis states.
At the large $J_2$ limit, $H_{\rm plane}$ can be
mapped to the well known spin-${1\over 2}$ single layer
Heisenberg antiferromagnet described by an effective Hamiltonian 
$H_{\rm eff}$, so we can
take $H_{\rm rung}$ and 
the $z$-component of  $H_{\rm eff}$
as the unperturbed Hamiltonian, and the rest as the perturbation.
The first Ising expansion works best in the
small $J_2/J_1$ region,
while the second one works best in the  
large  $J_2/J_1$ region.
Series are calculated for the ground-state energy, staggered magnetization, 
and triplet excitation spectra.
Extrapolating these series to the isotropic limit,
we find the system has long range N\'eel order and gapless excitation
for all ratios of interlayer to intralayer  couplings, 
as for the  square lattice. In Section III,
the 4 and 6-layer systems are studied by 
means of expansion in the interlayer
coupling $J_2$, and the transition points are located.
The last section is devoted to a summary. 

\section{System with odd number of coupled planes}
For $n=3$, the eigenstates for $h_{\rm rung}$  
 consist of


1) one $S_{\rm tot}=1/2$ doublet with energy 
$E_0^{\rm rung} =-1$ and eigenstates
\begin{eqnarray}
&S^z_{\rm tot}=-{1\over 2}: \quad & \vert ({1\over 2} , -{1\over 2},-1) \rangle ={1\over \sqrt{6}} (\vert \downarrow\downarrow\uparrow \rangle +
\vert \uparrow\downarrow\downarrow\rangle - 2 \vert \downarrow\uparrow\downarrow\rangle ) \nonumber \\ & &\\
&S^z_{\rm tot}={1\over 2}: \quad & \vert ({1\over 2} , {1\over 2},-1) \rangle = {1\over \sqrt{6}} (\vert \uparrow\uparrow\downarrow \rangle +
\vert \downarrow\uparrow\uparrow\rangle - 2 \vert \uparrow\downarrow\uparrow\rangle ) \nonumber 
\end{eqnarray}
where the arrows represent, from left to right, the $z$-components of
$S_{\alpha}$ ($\alpha=1,2,\cdots ,n$), and
the eigen states are denoted 
by $(S_{\rm tot}, S^z_{\rm tot}, E)$.
$S_{\rm tot}$, $S^z_{\rm tot}$ and $E$ are the total spin, 
the $z$-component of the total spin,
and its corresponding eigenvalue, respectively.
This is the ground state of $h_{\rm rung}$.


2) another $S_{\rm tot}=1/2$ doublet with energy $E_1^{\rm rung}=0$ and eigenstates
\begin{eqnarray}
&S^z_{\rm tot}=-{1\over 2}: \quad & \vert ({1\over 2} , -{1\over 2},0) \rangle = {1\over \sqrt{2}} (\vert \downarrow\downarrow\uparrow \rangle -
\vert \uparrow\downarrow\downarrow\rangle ) \nonumber \\ & &\\
&S^z_{\rm tot}={1\over 2}: \quad & \vert ({1\over 2} , {1\over 2},0) \rangle =  {1\over \sqrt{2}} (\vert \downarrow\uparrow\uparrow\rangle -
\vert \uparrow\uparrow\downarrow \rangle  ) \nonumber 
\end{eqnarray}

3) one $S_{\rm tot}={3\over 2}$ quartet with energy 
$E_2^{\rm rung}=1/2$ and eigenstates
\begin{eqnarray}
&S^z_{\rm tot}=-{3\over 2}: \quad &  \vert ({3\over 2},
-{3\over 2},{1\over 2}) \rangle = \vert \downarrow\downarrow\downarrow \rangle  \nonumber \\ 
&S^z_{\rm tot}=-{1\over 2}: \quad &  \vert ({3\over 2},
-{1\over 2},{1\over 2}) \rangle ={1\over \sqrt{3}} (\vert \downarrow\downarrow\uparrow \rangle + \vert \downarrow\uparrow\downarrow\rangle  
\vert \uparrow\downarrow\downarrow\rangle  ) \nonumber \\ & & \\
&S^z_{\rm tot}={1\over 2}: \quad & \vert ({3\over 2},{1\over 2} ,
{1\over 2}) \rangle ={1\over \sqrt{3}} (\vert \downarrow\uparrow\uparrow \rangle + \vert \uparrow\downarrow\uparrow\rangle  
\vert \uparrow\uparrow\downarrow\rangle  ) \nonumber \\
&S^z_{\rm tot}={3\over 2}: \quad & \vert ({3\over 2}, {3\over 2}, {1\over 2}) \rangle =\vert \uparrow\uparrow\uparrow \rangle  \nonumber 
\end{eqnarray}

So in the limit of $J_1/J_2=0$, the ground state of the 
whole system is the direct product of the ground states
of each rung, which are degenerate with $S={1\over 2}$
and $S^z = \pm {1\over 2}$, and so it is $2^N$-fold
degenerate ($N$ is no. of sites). 
A finite value of $J_1$ lifts this 
degeneracy (to 2). In this $2^N$-dimensional subspace, 
one can define an effective Hamiltonian
$H_{\rm eff}$ which includes all interactions 
$H_{\rm plane}$. To first order in $J_1/J_2$ 
we obtain\cite{rei94} 
\begin{equation}
H_{\rm eff} = J_{\rm eff} \sum_{\langle ij\rangle} 
{\bf S}^{\rm eff}_i \cdot {\bf S}^{\rm eff}_j \label{heff}
\end{equation}
with effective coupling 
\begin{equation}
J_{\rm eff}=J_1
\end{equation}
Note that $J_{\rm eff}$
for Heisenberg ladder systems has been computed by
N.  Hatano and Y. Nishiyama\cite{rei94},
and their results can also be applied to the multi-plane 
Heisenberg antiferromagnet discussed here.


Therefore the ground state energy per site to 
first order in $J_1/J_2$  
will be
\begin{equation}
E_0/N =(J_2 E_0^{\rm rung} + E_0^{\rm sq} J_{\rm eff})/n \label{e0j2}
\end{equation}
where $E_0^{\rm sq}=-0.6693$ is the ground state energy per site for the 
square lattice\cite{zwh_sq}.

To find out the staggered magnetization 
at large $J_2$ limit, 
one needs to look at the results of the operator
$S_1^z - S_2^z + S_3^z$ on the states $\vert ({1\over 2}, -{1\over 2}, -1) \rangle $
and $\vert ({1\over 2}, {1\over 2}, -1) \rangle $:
\begin{eqnarray}
(S_1^z - S_2^z + S_3^z )\vert ({1\over 2}, -{1\over 2}, -1) \rangle &=& -{5\over 6} \vert ({1\over 2}, -{1\over 2}, -1) \rangle  + 0.942809 \vert ({3\over 2}, -{1\over 2}, {1\over 2}) \rangle  \nonumber \\ && \\
(S_1^z - S_2^z + S_3^z )\vert ({1\over 2}, {1\over 2}, -1) \rangle &=& {5\over 6} \vert ({1\over 2}, {1\over 2}, -1) \rangle - 0.942809 \vert ({3\over 2}, {1\over 2}, {1\over 2}) \rangle \nonumber
\end{eqnarray}
If we only keep the states which are the ground states of $h_{\rm rung}$,
and compare the above relations with that for the 
single layer square lattice,
we can get the staggered magnetization at the large $J_2$ limit as
\begin{equation}
M \equiv  {1\over N} \sum_{i,\alpha} \langle (-1)^{i+\alpha} S_{i,\alpha}^z \rangle_0 
= {5 M^{\rm sq}\over 3n} = 0.1706(6)
\end{equation}
where $M^{\rm sq}=0.307(1)$ is the staggered magnetization 
for the single layer square lattice.\cite{zwh_sq}

For other odd $n$ values, 
the ground states of $h_{\rm rung}$  are also 
two spin-${1\over 2}$ doublet states.
In Table I, we listed the ground state energy 
$E_0^{\rm rung}$
 of $h_{\rm rung}$,
the effective coupling $J_{\rm eff}$ and 
the staggered magnetization $M$ at the large $J_2$ limit
for $n=3,5,7,9,11$, 
and these are also shown
in Figure 1, where 
we can see that as $n$ increases, 
$J_{\rm eff}/J_1$ is generally 
close to unity but increases monotonically,
while $M/M^{\rm sq}$ decreases about as $n^{-1/2}$.
At the limit of $n\to\infty$, we expect that 
$J_{\rm eff}/J_1$ approaches
a constant\cite{rei94}, while 
$M/M^{\rm sq}$ decreases to zero.


Since at both large and small $y$ limits, the system can be reduced to 
a single plane Heisenberg antiferromagnet, we expect that unlike the
bilayer Heisenberg antiferromagnets, there is no 
transition at any value of $y$.

To study the properties of these systems, we develop following two 
different Ising expansions. 

The first one is the usual naive Ising expansion where we take 
 the $z$-component of the Hamiltonian as the unperturbed
Hamiltonian $H_0$, and the rest of it as the perturbation $V$:
we write the Hamiltonian  for the Heisenberg-Ising model as:
\begin{equation}
H/J_1 = H_0 + x V  \label{Hising}
\end{equation}
where 
\begin{eqnarray}
H_0 &= & \sum_{\alpha=1}^n \sum_{\langle i,j\rangle} 
S_{\alpha,i}^z S_{\alpha,j}^z
+ y \sum_{\alpha=1}^{n-1}  \sum_{i}  S_{\alpha,i}^z S_{\alpha+1,i}^z  + 
 t \sum_{\alpha,i} \epsilon_{\alpha,i} S_{\alpha,i}^z \nonumber \\ && \\
V &= & \sum_{\alpha=1}^{n} \sum_{\langle i,j\rangle} 
 ( S_{\alpha,i}^x S_{\alpha,j}^x + S_{\alpha,i}^y S_{\alpha,j}^y )  + 
y \sum_{\alpha=1}^{n-1} \sum_{i} ( S_{\alpha,i}^x S_{\alpha+1,i}^x + S_{\alpha,i}^y S_{\alpha+1,i}^y ) 
- t \sum_{\alpha,i} \epsilon_{\alpha,i} S_{\alpha,i}^z \nonumber
\end{eqnarray}
and $\epsilon_{\alpha,i}=\pm 1 $ on the two sublattices. The last term
in both $H_0$ and $V$ is a local staggered field term, which can be
included to improve convergence.
The limits $x=0$ and $x=1$ correspond to the Ising model and
the isotropic Heisenberg model respectively.
The operator $H_0$ is taken as the unperturbed
Hamiltonian, with the unperturbed ground state being the 
usual N\'{e}el state.
The operator $V$ is treated as  a perturbation.
It flips a pair of spins on neighbouring sites. 
The linked-cluster expansion method has been previously reviewed
in several articles\cite{he90,gel90,gelmk}, and will not be repeated here.

The Ising series have been calculated for 
the ground state energy per site $E_0/N$,
the staggered magnetization $M$,  
and the lowest lying triplet excitation spectrum $\Delta (k_x, k_y)$
for several ratios of couplings and (simultaneously) for several values of $t$
 up to order $x^{11}$ for $n=3$ or order $x^9$ for $n=5$.
The triplet excitation spectrum is computed for $n=3$ only.
The resulting series for $E_0/N$ and $M$ for 
$y=0.5, 1, 2$  are listed in Tables II,
and the other series  are 
available on request. 

There are three bands of the spin-wave dispersion 
for 3-layer systems. In the
Ising limit, 2 bands have initial excitations
located in the side planes, and the third band has it
in the middle plane. 
From the series  one can see
that all three bands have  the following symmetry:
\begin{equation}
\Delta (k_x, k_y) = \Delta (k_y, k_x) = \Delta (\pi-k_x, \pi-k_y)
\end{equation}

Since at the large $J_2$ limit the system can be mapped to 
the well known spin-${1\over 2}$ single layer
Heisenberg antiferromagnet with effective Hamiltonian $H_{\rm eff}$
defined in Eq. (\ref{heff}), we can develop,
in the basis of the eigenstates of $h_{\rm rung}$, 
another Ising
expansion, the so called ``modified Ising expansion'', 
where we take $H_{\rm rung}$ and  the $z$-component of $H_{\rm eff}$
%
as the unperturbed Hamiltonian, and the rest as the 
perturbation.
That is, we write the Hamiltonian for the Heisenberg-Ising model as:
\begin{equation}
H = H'_0 + x V'  \label{Hising2}
\end{equation}
where 
\begin{eqnarray}
H'_0 &= & J_2 H_{\rm rung} + J_{\rm eff} \sum_{\langle i,j\rangle} 
S_{i}^{{\rm eff},z} S_{j}^{{\rm eff},z}
+ t \sum_{i} \epsilon_{i} S_{i}^{{\rm tot},z} \nonumber \\ && \\
V' &= & J_1 H_{\rm plane}
 - J_{\rm eff} \sum_{\langle i,j\rangle} 
S_{i}^{{\rm eff},z} S_{j}^{{\rm eff},z}
- t \sum_{i} \epsilon_{i} S_{i}^{{\rm tot},z} \nonumber
\end{eqnarray}
$S_{i}^{{\rm tot},z}$ is $z$-component of 
the total spin for $i$th rung, and
 again the last term
in both $H'_0$ and $V'$ is a local staggered field term, which can be
included to improve convergence.
The limits $x=0$ and $x=1$ correspond to the Ising model and
the isotropic Heisenberg model respectively.
The operator $H'_0$ is taken as the unperturbed
Hamiltonian, with the unperturbed ground state being the 
usual N\'{e}el state.
The operator $V'$ is treated as  a perturbation.
The detailed eigenvalues and eigenstates of $h_{\rm rung}$ 
and also the matrix elements $V'$ under 
the eigenstates of $h_{\rm rung}$ for both $n=3$
and $n=5$ are available from the author.

Here the Ising series have been calculated for 
the ground state energy per site $E_0/N$, and
the staggered magnetization $M$
for several ratios of couplings and (simultaneously) for several values of $t$
 up to order $x^6$ for $n=3$ or order $x^4$ for $n=5$.
The resulting series for $E_0/N$ and $M$ for 
$y=5, 10, 20$  are listed in Tables II,
and the other series  are 
available on request. 

Having obtained the series for the above two Ising 
expansions we  try to extrapolate
the series to the isotropic point ($x=1$).
For this purpose, we first transform the series to a new variable 
\begin{equation}
\delta = 1- (1-x)^{1/2}
\end{equation}
to remove the singularity at $x=1$ predicted by the spin-wave theory. This 
was first proposed by Huse\cite{hus} and was also
used in our earlier work on the square lattice case\cite{zwh_sq}.
We then use both  integrated first-order inhomogeneous
differential approximants\cite{gut} and  naive Pad\'{e} approximants to
extrapolate the series to the isotropic point $\delta=1$  ($x=1$).
The results for the ground state energy per site $E_0/N$,
and the staggered magnetization $M$
are shown in Figs. 2-3. 
The results for $J_2=0$ presented in the figures 
are taken from our earlier work on the single-layer 
square lattice\cite{zwh_sq}, while the results for
$J_1/J_2=0$ are taken from the mapping of 
the system on to the
single layer system.
The asymptotic behaviour, Eq. (\ref{e0j2}), of the ground state energy 
is also shown in Fig. 2 as dotted line, and
they match on to those Ising expansions very well.
The results from two Ising expansion also match
each other very well in certain regions of coupling.
We note that $M$ first increases 
for small $J_2/J_1$,
passes through a maximum at about $J_2/J_1\simeq 1$, and 
then decreases  for
larger values of $J_2/J_1$.
The reason that in the case of small $J_2/J_1$,
the interlayer coupling enhances the antiferromagnetic long-range order
is that the system acquires a weak three 
dimensionality and 
quantum fluctuations are suppressed. 
The spin-wave theory for the bilayer Heisenberg 
antiferromagnet predicts that $M$ should be 
proportional to $(J_2/J_1)^{1/2}$ in the small 
$J_2/J_1$ region, we can expect this will 
happen for any multi-layer system,
and so in Figs. 2-3 we show $E_0$ and $M$ as a 
function of $(J_2/J_1)^{1/2}/[J_2/J_1)^{1/2}+1]$
to exhibit this behaviour.
From Fig. 3, we can see that in the 
small $J_2/J_1$ region, the 5-layer system has slightly
stronger Ne\'el order than the 3-layer system, 
while in the large $J_2/J_1$ region,
the 3-layer system has stronger Ne\'el order.

Figs. 4-6 show the three bands of triplet excitation
spectra of the 
3-layer system obtained from the naive Ising expansion. 
From these graphs, we can see that all of 
the dispersion relations have
 a minimum located at $(k_x,k_y)=(0,0)$ (or $(\pi,\pi)$ by symmetry),
where two of these three bands have a definite gap as long as $J_2\neq 0$,
while the third (the outer symmetry  band) is consistent with 
a gapless spectrum for all values of $J_2$.

\section{System with even number of coupled planes}
For even $n$, the ground state of $h_{\rm rung}$ 
is a $S=0$ singlet. For example, the eigenstates of $h_{\rm rung}$ 
for $n=4$ consist of

%

1) one $S_{\rm tot}=0$ singlet with energy $E_0^{\rm rung}
=(-3-2 \sqrt{3})/4$ and eigenstate
\begin{equation}
\vert (0,0,E_0^{\rm rung})\rangle =  c_1 (\vert \downarrow\downarrow\uparrow\uparrow\rangle + \vert \uparrow\uparrow\downarrow\downarrow\rangle )
-c_2 (\vert \downarrow\uparrow\downarrow\uparrow\rangle + \vert \uparrow\downarrow\uparrow\downarrow\rangle )
+c_3 (\vert \uparrow\downarrow\downarrow\uparrow\rangle + \vert \downarrow\uparrow\uparrow\downarrow\rangle )
\end{equation}
where
\begin{equation}
c_1 = {{3 + 2\,{\sqrt{3}}}\over {6\,{\sqrt{26 + 15\,{\sqrt{3}}}}}} 
,\quad  c_2 = {{ 12 + 7\,{\sqrt{3}} }\over
     {6\,{\sqrt{26 + 15\,{\sqrt{3}}}}}} 
,\quad  c_3 = {{9 + 5\,{\sqrt{3}}}\over {6\,{\sqrt{26 + 15\,{\sqrt{3}}}}}}
\end{equation}
This is the ground state of $H_{\rm rung}$.

2) one $S_{\rm tot}=1$ triplet with energy $E_1^{\rm rung}=(-1-2\sqrt{2})/4$ and eigenstates
\begin{eqnarray}
&S^z_{\rm tot}=-1: \quad & \vert (1,-1,E_1^{\rm rung})\rangle = c_4 (\vert \downarrow\downarrow\downarrow\uparrow\rangle-\vert\uparrow\downarrow\downarrow\downarrow\rangle)
+c_5 (\vert\downarrow\uparrow\downarrow\downarrow\rangle-\vert\downarrow\downarrow\uparrow\downarrow\rangle) \nonumber \\
&S^z_{\rm tot}=0: \quad & \vert (1,0,E_1^{\rm rung})\rangle = c_4 (\vert\downarrow\downarrow\uparrow\uparrow\rangle-\vert\uparrow\uparrow\downarrow\downarrow\rangle)
+c_5(\vert\uparrow\downarrow\uparrow\downarrow\rangle-\vert\downarrow\uparrow\downarrow\uparrow\rangle)
 \\
&S^z_{\rm tot}=1: \quad &  \vert (1,1,E_1^{\rm rung})\rangle = c_4 (\vert\downarrow\uparrow\uparrow\uparrow\rangle-\vert\uparrow\uparrow\uparrow\downarrow\rangle)
+ c_5 (\vert\uparrow\uparrow\downarrow\uparrow\rangle-\vert\uparrow\downarrow\uparrow\uparrow\rangle) \nonumber 
\end{eqnarray}
where
\begin{equation}
c_4 = {{1 + {\sqrt{2}}}\over {2\,{\sqrt{10 + 7\,{\sqrt{2}}}}}} 
, \quad 
c_5 = {{3 + 2\,{\sqrt{2}}}\over {2\,{\sqrt{10 + 7\,{\sqrt{2}}}}}} 
\end{equation}

3) another $S_{\rm tot}=1$ triplet with energy $E_2^{\rm rung}=-1/4$ and eigenstates:
\begin{eqnarray}
&S^z_{\rm tot}=-1: \quad & \vert (1,-1,E_2^{\rm rung})\rangle = {1\over 2}
(\vert\downarrow\downarrow\downarrow\uparrow\rangle+\vert\uparrow\downarrow\downarrow\downarrow\rangle-\vert\downarrow\downarrow\uparrow\downarrow\rangle-\vert\downarrow\uparrow\downarrow\downarrow\rangle)\nonumber \\
&S^z_{\rm tot}=0:  \quad &\vert (1,0,E_2^{\rm rung})\rangle = {1\over \sqrt{2}}(\vert\uparrow\downarrow\downarrow\uparrow\rangle-\vert\downarrow\uparrow\uparrow\downarrow\rangle)  \\
&S^z_{\rm tot}=1:  \quad &\vert (1,1,E_2^{\rm rung})\rangle = {1\over 2}(\vert\downarrow\uparrow\uparrow\uparrow\rangle+\vert\uparrow\uparrow\uparrow\downarrow\rangle-\vert\uparrow\downarrow\uparrow\uparrow\rangle-\vert\uparrow\uparrow\downarrow\uparrow\rangle)\nonumber 
\end{eqnarray}

4) another $S_{\rm tot}=0$ singlet with energy $E_3^{\rm rung}=(-3+2 \sqrt{3})/4$ and eigenstate
\begin{equation}
\vert (0,0,E_3^{\rm rung})\rangle = c_2 (\vert\downarrow\downarrow\uparrow\uparrow\rangle+\vert\uparrow\uparrow\downarrow\downarrow\rangle)- c_1 (\vert\downarrow\uparrow\downarrow\uparrow\rangle+\vert\uparrow\downarrow\uparrow\downarrow\rangle)
- c_3 (\vert\uparrow\downarrow\downarrow\uparrow\rangle+\vert\downarrow\uparrow\uparrow\downarrow\rangle)
\end{equation}

5) another $S_{\rm tot}=1$ triplet with energy $E_4^{\rm rung}=(-1+2 \sqrt{2})/4$ and eigenstates:
\begin{eqnarray}
&S^z_{\rm tot}=-1: \quad & \vert (1,-1,E_4^{\rm rung})\rangle = c_5 (\vert\downarrow\downarrow\downarrow\uparrow\rangle-\vert\uparrow\downarrow\downarrow\downarrow\rangle)
+ c_4 (\vert\downarrow\downarrow\uparrow\downarrow\rangle-\vert\downarrow\uparrow\downarrow\downarrow\rangle) \nonumber \\
&S^z_{\rm tot}=0:  \quad &  \vert (1,0,E_4^{\rm rung})\rangle = c_5 (\vert\downarrow\downarrow\downarrow\uparrow\rangle-\vert\uparrow\downarrow\downarrow\downarrow\rangle)
+ c_4 (\vert\downarrow\downarrow\uparrow\downarrow\rangle-\vert\downarrow\uparrow\downarrow\downarrow\rangle)  \\
&S^z_{\rm tot}=1:  \quad & \vert (1,1,E_4^{\rm rung})\rangle =   c_5 (\vert\downarrow\uparrow\uparrow\uparrow\rangle-\vert\uparrow\uparrow\uparrow\downarrow\rangle)
+ c_4 (\vert\uparrow\downarrow\uparrow\uparrow\rangle-\vert\uparrow\uparrow\downarrow\uparrow\rangle) \nonumber 
\end{eqnarray}

6) one $S_{\rm tot}=2$ quintuplet with energy $E_5^{\rm rung}=3/4$ and eigenstates
\begin{eqnarray}
&S^z_{\rm tot}=-2:  \quad & \vert (2,-2,E_5^{\rm rung})\rangle = \vert\downarrow\downarrow\downarrow\downarrow\rangle \nonumber \\
&S^z_{\rm tot}=-1:  \quad & \vert (2,-1,E_5^{\rm rung})\rangle ={1\over 2}(\vert\downarrow\downarrow\downarrow\uparrow\rangle
+\vert\downarrow\downarrow\uparrow\downarrow\rangle+\vert\downarrow\uparrow\downarrow\downarrow\rangle+\vert\uparrow\downarrow\downarrow\downarrow\rangle) \nonumber \\
&S^z_{\rm tot}=0:  \quad & \vert (2,0,E_5^{\rm rung})\rangle = {1\over \sqrt{6}} (\vert\downarrow\downarrow\uparrow\uparrow\rangle+\vert\downarrow\uparrow\downarrow\uparrow\rangle
+\vert\uparrow\downarrow\downarrow\uparrow\rangle+\vert\downarrow\uparrow\uparrow\downarrow\rangle+\vert\uparrow\downarrow\uparrow\downarrow\rangle+\vert\uparrow\uparrow\downarrow\downarrow\rangle)  \\
&S^z_{\rm tot}=1:  \quad & \vert (2,1,E_5^{\rm rung})\rangle ={1\over 2}(\vert\downarrow\uparrow\uparrow\uparrow\rangle+\vert\uparrow\downarrow\uparrow\uparrow\rangle+\vert\uparrow\uparrow\downarrow\uparrow\rangle+
\vert\uparrow\uparrow\uparrow\downarrow\rangle) \nonumber \\
&S^z_{\rm tot}=2:  \quad & \vert (2,2,E_5^{\rm rung})\rangle =\vert\uparrow\uparrow\uparrow\uparrow\rangle \nonumber 
\end{eqnarray}

So at the limit of $J_1/J_2=0$, the ground state of the
whole system is the direct product of the ground state
of each rung. Unlike the case for odd value of $n$,
there is not degeneracy in this ground state, and 
the system has a definite excitation gap. 

The operator $S_1^z - S_2^z + S_3^z - S_4^z$ on the ground state 
$\vert (0,0,E_0^{\rm rung}) \rangle $ gives
\begin{equation}
(S_1^z - S_2^z + S_3^z - S_4^z )\vert (0,0,E_0^{\rm rung}) \rangle 
= -1.45728 \vert (1,0,E_1^{\rm rung})\rangle + 0.6036258 
\vert (1,0,E_4^{\rm rung})\rangle 
\end{equation}
Again unlike the case for odd $n$, the right hand side
of the above equation does not contain the ground
state $\vert (0,0,E_0^{\rm rung}) \rangle $,
so in the large $J_2$ limit,
the staggered magnetization $M$ is zero, that is,
the system is in a quantum disordered spin liquid state.
So we expect there is a transition
separating the gapless Ne\'el phase and gapped 
quantum disordered spin liquid
phase at a certain critical ratio of $J_2/J_1$,
and we expect that this is true for any even number of layers.


To locate the transition point, we can develop the 
expansions around the large interlayer coupling limit,
that is, we can construct an expansion in $1/y$ 
by treating the operator $ H_{\rm rung}$ as the 
unperturbed Hamiltonian and
operator $H_{\rm plane}$ as a perturbation.
%
We have developed the expansion for the $T=0$ ground state energy
per site $E_0/N$,  the antiferromagnetic susceptibility $\chi$,
and the lowest lying triplet excitation spectrum $\Delta (k_x, k_y)$
up to a certain order (see Table III) for $n=4$ and 6.
The resulting power series in $1/y$ for $E_0/N$ 
and $\chi$ are presented in Table III.
The series for the minimum energy gap $m=\Delta (\pi, \pi)$ is also listed
in Table III. The series for the excitations spectra
$\Delta (k_x, k_y)$ are available on request. 

To determine the critical point, we construct  Dlog Pad\'{e}
approximants\cite{gut} to the $\chi$ and $m$ series.
Since the transition should lie in
the universality class of the classical $d=3$ Heisenberg model
(our analysis also supports this), we expect that the critical
index for $\chi$ and $m$
should be approximately  1.40 and 0.71, respectively. The
exponent-biased Dlog Pad\'{e} approximants
give the critical point $(J_1/J_2)_c=0.213(5)$ for $n=4$, and
$(J_1/J_2)_c=0.145(10)$ for $n=6$.
The critical points $(J_1/J_2)_c$
as function of $1/n$ are shown in Fig. 7 
(the results for $n=2$ are taken
from a previous calculation\cite{zwh_bilayer}),
where we can see they can be well fitted by
 $(J_1/J_2)_c=0.913/n-0.250/n^2$. So 
 $(J_1/J_2)_c$ should be 
proportional to $1/n$ at the large $n$ limit.

The triplet excitation spectra $\Delta(k_x,k_y)$
at the critical point for $n=4,6$ are
illustrated in Fig. 8(a): the results for $n=2$ are 
also displayed here for ease of comparison.
We can see, as in the case of the bilayer
Heisenberg antiferromagnet\cite{zwh_bilayer},
the direct sum to the series is indeed consistent
with the Pad\'{e} approximants that one can construct.  
The spectra have their maximum value 
$\Delta(0,0)$ at $(k_x,k_y)=(0,0)$.
In Fig. 8(b), we plot the function for
$\Delta(k_x,k_y)/\Delta(0,0)$ along high-symmetry cuts
through the Brillouin zone, and we can see
this function appears to be 
an universal function,
independent of the number of layers $n$. We have no independent
theory to explain it at present,
but argue this may be due to the fact
they belong to the same universality class.

\section{Conclusions}
We have presented and analyzed  
various perturbation series expansions 
for  multi-layer, $S={1\over 2}$, square lattice 
Heisenberg model with up to
6 coupled planes. Like Heisenberg ladder system, the
 Heisenberg antiferromagnet on 
an odd number of coupled
planes shows  a fundamentally different behavior
from that for an even number of coupled planes.

For an odd number of layers, the system can be mapped  
to the well-known single layer Heisenberg
antiferromagnet at the large $J_2$ limit, 
and so the system should have 
Ne\'el order for all ratios of interlayer to intralayer  
couplings:
that is, it is similar to the case of the square lattice.
The staggered magnetization $M$ at the
large $J_2/J_1$ limit is computed up to 11-layers, 
and we found $M$ should decrease 
to zero as $n^{-1/2}$
in the large $n$ limit.
For the systems with 3 and 5-layers, 
two Ising expansions have been developed, and
series are calculated for the ground-state energy, the
staggered magnetization, 
and the triplet excitation spectra.
Extrapolating these series to the isotropic limit,
we find the system has long range N\'eel order 
and gapless excitations for all ratios of 
interlayer to intralayer couplings.

For an even numbers of layers, expansions in the interlayer coupling
have been developed  for both 4 and 6 layers. 
The systems turn out to have
a second order transition separating a gapless 
Ne\'el phase and a gapped quantum disordered spin
liquid phase, where the critical point
is determined to be $(J_1/J_2)_c=0.213(5)$ for $n=4$, and
$(J_1/J_2)_c=0.145(10)$ for $n=6$, and this critical point
should be proportional to $1/n$ it the large $n$ limit.
The triplet excitation
spectra are also computed, and at the critical point, 
it turns out that the
excitation spectrum, after normalization by its
maximum value, appears to be an universal function,
independent of the number of layers.
This may be due to the fact that they belong to the 
same universality class.

\acknowledgments
This work forms part of a research project supported by a grant 
from the Australian Research Council. 
I would like to thank Profs.
J. Oitmaa and C.J. Hamer
for a number of valuble discussions and suggestions.


\begin{figure}[htb]
\caption{The effective coupling $J_{\rm eff}$ and the staggered 
Magnetization $M$ for odd $n$ at the limit
$J_2/J_1\to\infty$ as function of $n^{-1/2}$, and the solid line is
the least square polynomial fit 
$M/M^{\rm sq}=1.125n^{-1/2}-0.532n^{-1}+0.431n^{-3/2}$.
}
\label{fig:fig1}
\end{figure}

\begin{figure}[htb]
\caption{The rescaled ground-state energy per 
site $E_0/N$ for 3 and 5-layers as 
function of $(J_2/J_1)^{1/2}/[(J_2/J_1)^{1/2}+1]$.
The solid (open) points with error bars are the estimates from
the naive (modified) Ising expansions, 
and the dotted curve at large $(J_2/J_1)^{1/2}$ region 
is the  asymptotic behaviour, i.e., Eq. (\ref{e0j2}).
}
\label{fig:fig2}
\end{figure}

\begin{figure}[htb]
\caption{The staggered magnetization $M$ for 3 and 5-layers
{\it versus} $(J_2/J_1)^{1/2}/[(J_2/J_1)^{1/2}+1]$.
The solid (open) points with error bars are the estimates from
the naive (modified) Ising expansions.
}
\label{fig:fig3}
\end{figure}

\begin{figure}[htb]
\caption{Plot of 
symmetric spin-triplet excitation (in side planes) spectrum 
$\Delta(k_x, k_y)$ 
(derived from the naive Ising expansions) along high-symmetry
cuts through the Brillouin zone for the three-layer system with coupling ratios
$J_2/J_1=0, 0.5, 1$ (shown in
the figure from  the bottom to the top  respectively).}
\label{fig:fig4}
\end{figure}

\begin{figure}[htb]
\caption{Plot of 
antisymmetric spin-triplet excitation (in side planes) spectrum 
$\Delta(k_x, k_y)$
(derived from the naive Ising expansions) along high-symmetry
cuts through the Brillouin zone for the three-layer system with coupling ratios
$J_2/J_1=0, 0.5, 1, 2, 4$ (shown in
the figure from  the bottom to the top  respectively).}
\label{fig:fig5}
\end{figure}

\begin{figure}[htb]
\caption{Plot of
 spin-triplet excitation (in middle plane) spectrum 
$\Delta(k_x, k_y)$
(derived from the naive Ising expansions) along high-symmetry
cuts through the Brillouin zone for the three-layer system with coupling ratios
$J_2/J_1=0, 0.1, 0.25, 0.5, 1$ (shown in
the figure from  the bottom to the top  respectively).}
\label{fig:fig6}
\end{figure}

\begin{figure}[htb]
\caption{Plot of the critical points $(J_1/J_2)_c$ 
{\it versus} $1/n$ for even number of coupled planes.
The solid points with error bars are the estimates from
expansions about interlayer coupling, and the solid line is
the least square polynomial fit $(J_1/J_2)_c=0.913/n-0.250/n^2$.
\label{fig:fig7}
}
\end{figure}

\begin{figure}[htb]
\caption{Plot of the triplet excitation spectrum
$\Delta(k_x,k_y)$ (a) and spectrum after normalization by its
maximum value (b) along high-symmetry
cuts through the Brillouin zone for the $n=2,4,6$ systems 
at the transition point,
the lines are the estimates by direct sum to the dimer series,
and the points 
are the estimates of the Pad\'{e} approximants to the series.
}
\label{fig:fig8}
\end{figure}

\begin{table}
\squeezetable
\setdec 0.00000000000
\caption{the ground state energy $E_0^{\rm rung}$ of $h_{\rm rung}$,
the effective coupling $J_{\rm eff}$ and 
the staggered magnetization $M$ in the limit $J_2/J_1\to \infty$
for odd $n$ up to $n=11$.
}
 \label{tab1}
\begin{tabular}{rrrrr}
 \multicolumn{1}{c}{n} &\multicolumn{1}{c}{$E_0^{\rm rung}/n$}
&\multicolumn{1}{c}{$J_{\rm eff}/J_1$} 
&\multicolumn{1}{c}{$M/M^{\rm sq}$}
 \\
\tableline
 3 & $-$0.33333333 & 1             & 0.555556 \\
 5 & $-$0.38557725 & 1.016938829   & 0.435575 \\
 7 & $-$0.40517710 & 1.034391333   & 0.372626 \\
 9 & $-$0.41514686 & 1.050406957   & 0.331978 \\ 
11 & $-$0.42109939 & 1.064819159   & 0.302836 \\
\end{tabular}
\end{table}

\widetext
\begin{table}
\squeezetable
\setdec 0.0000000000
\caption{Series coefficients  of naive and modified 
Ising expansions of three layer system for  the
ground-state energy per site $E_0/(NJ_1)$ and the staggered magnetization $M$,
 for $J_2/J_1=0.5,1,2$ (for naive Ising expansion) $5,10,20$ 
(for modified Ising expansion)
and $t=0$.
Nonzero coefficients $x^i$
up to order $i=10$ for naive Ising expansion or
to order $i=6$ for modified Ising expansion are listed.}
 \label{tab2}
\begin{tabular}{r|rrr|rrr}
\multicolumn{1}{c|}{$i$}&\multicolumn{3}{c|}{$E_0/(J_1N)$}&\multicolumn{3}{c}{$M$}\\
\tableline
\multicolumn{1}{c|}{}&\multicolumn{6}{c}{Naive Ising expansion} \\
\multicolumn{1}{c|}{} 
&\multicolumn{1}{c}{$J_2/J_1=0.5$}&\multicolumn{1}{c}{$J_2/J_1=1$} 
&\multicolumn{1}{c|}{$J_2/J_1=2$}
&\multicolumn{1}{c}{$J_2/J_1=0.5$} &\multicolumn{1}{c}{$J_2/J_1=1$} 
&\multicolumn{1}{c}{$J_2/J_1=2$}   \\
  0 &\dec $-$3.33333333$\times 10^{-1}$ &\dec $-$4.16666667$\times 10^{-1}$ &\dec $-$5.83333333$\times 10^{-1}$ &\dec  5.00000000$\times 10^{-1}$ &\dec  5.00000000$\times 10^{-1}$ &\dec  5.00000000$\times 10^{-1}$ \\
  2 &\dec $-$1.46708683$\times 10^{-1}$ &\dec $-$1.53703704$\times 10^{-1}$ &\dec $-$2.23809524$\times 10^{-1}$ &\dec $-$7.98687122$\times 10^{-2}$ &\dec $-$7.14609053$\times 10^{-2}$ &\dec $-$8.68027211$\times 10^{-2}$ \\
  4 &\dec  1.10044693$\times 10^{-3}$ &\dec  5.08576750$\times 10^{-4}$ &\dec  1.36582616$\times 10^{-3}$ &\dec $-$6.95556119$\times 10^{-3}$ &\dec $-$5.95673893$\times 10^{-3}$ &\dec $-$8.37837732$\times 10^{-3}$ \\
  6 &\dec $-$1.39063516$\times 10^{-3}$ &\dec $-$1.97000255$\times 10^{-3}$ &\dec $-$2.95007447$\times 10^{-3}$ &\dec $-$4.80430415$\times 10^{-3}$ &\dec $-$5.53318319$\times 10^{-3}$ &\dec $-$7.46816918$\times 10^{-3}$ \\
  8 &\dec $-$5.11831870$\times 10^{-4}$ &\dec $-$6.87771224$\times 10^{-4}$ &\dec $-$1.22506627$\times 10^{-3}$ &\dec $-$3.03361100$\times 10^{-3}$ &\dec $-$3.35287068$\times 10^{-3}$ &\dec $-$5.27035727$\times 10^{-3}$ \\
 10 &\dec $-$2.75677029$\times 10^{-4}$ &\dec $-$4.09491760$\times 10^{-4}$ &\dec $-$6.02299585$\times 10^{-4}$ &\dec $-$2.01622805$\times 10^{-3}$ &\dec $-$2.42979325$\times 10^{-3}$ &\dec $-$3.42232873$\times 10^{-3}$ \\
\tableline
\multicolumn{1}{c|}{}&\multicolumn{6}{c}{Modified Ising expansion} \\
\multicolumn{1}{c|}{} 
&\multicolumn{1}{c}{$J_2/J_1=5$}&\multicolumn{1}{c}{$J_2/J_1=10$} 
&\multicolumn{1}{c|}{$J_2/J_1=20$}
&\multicolumn{1}{c}{$J_2/J_1=5$} &\multicolumn{1}{c}{$J_2/J_1=10$} 
&\multicolumn{1}{c}{$J_2/J_1=20$}   \\
  0 &\dec $-$1.83333333 &\dec $-$3.50000000 &\dec $-$6.83333333 &\dec  2.77777778$\times 10^{-1}$ &\dec  2.77777778$\times 10^{-1}$ &\dec  2.77777778$\times 10^{-1}$ \\
  1 &\dec  0.00000000 &\dec  0.00000000 &\dec  0.00000000 &\dec  6.97167756$\times 10^{-2}$ &\dec  3.70370370$\times 10^{-2}$ &\dec  1.91158901$\times 10^{-2}$ \\
  2 &\dec $-$1.21179736$\times 10^{-1}$ &\dec $-$9.07296550$\times 10^{-2}$ &\dec $-$7.38090200$\times 10^{-2}$ &\dec $-$4.36538214$\times 10^{-2}$ &\dec $-$5.67336122$\times 10^{-2}$ &\dec $-$6.04157972$\times 10^{-2}$ \\
  3 &\dec $-$1.41804384$\times 10^{-2}$ &\dec $-$3.48154003$\times 10^{-3}$ &\dec $-$6.42216094$\times 10^{-4}$ &\dec $-$1.35350826$\times 10^{-2}$ &\dec $-$7.15990062$\times 10^{-3}$ &\dec $-$3.64767685$\times 10^{-3}$ \\
  4 &\dec  4.43598485$\times 10^{-3}$ &\dec  3.25869698$\times 10^{-3}$ &\dec  1.90188695$\times 10^{-3}$ &\dec $-$5.47105059$\times 10^{-3}$ &\dec $-$4.79712404$\times 10^{-3}$ &\dec $-$6.78984443$\times 10^{-3}$ \\
  5 &\dec  5.30206875$\times 10^{-3}$ &\dec  1.33866469$\times 10^{-3}$ &\dec  3.36884895$\times 10^{-4}$ &\dec  9.81694291$\times 10^{-3}$ &\dec  2.83116612$\times 10^{-3}$ &\dec  5.48197309$\times 10^{-4}$ \\
  6 &\dec $-$1.30627826$\times 10^{-3}$ &\dec $-$1.28650672$\times 10^{-3}$ &\dec $-$8.82824083$\times 10^{-4}$ &\dec $-$3.28072052$\times 10^{-3}$ &\dec $-$5.39705234$\times 10^{-3}$ &\dec $-$5.29473180$\times 10^{-3}$ \\
\end{tabular}
\end{table}

\widetext
\begin{table}
\squeezetable
\setdec 0.0000000000
\caption{Series coefficients for expansions about interlayer coupling of the
ground-state energy per site $E_0/(NJ_2)$, the $(\pi,\pi)$ gap  $m/J_2$, and the
antiferromagnet susceptibility $\chi$
 for 4 and 6 layer systems.
Nonzero coefficients $(J_1/J_2)^i$ up to maximum order carried out are listed.}
 \label{tab3}
\begin{tabular}{r|rrr|rrr}
\multicolumn{1}{c|}{} &\multicolumn{3}{c|}{4 layers} &\multicolumn{3}{c}{6 layers} \\
\multicolumn{1}{c|}{i} &\multicolumn{1}{c}{$E_0/(NJ_2)$} &\multicolumn{1}{c}{$m/J_2$} 
&\multicolumn{1}{c|}{$\chi$} &\multicolumn{1}{c}{$E_0/(NJ_2)$} &\multicolumn{1}{c}{$m/J_2$} 
&\multicolumn{1}{c}{$\chi$} \\
\tableline
  0 &\dec $-$4.040063509$\times 10^{-1}$ &\dec  6.589186226$\times 10^{-1}$ &\dec  1.699358737 &\dec $-$4.155961890$\times 10^{-1}$ &\dec  4.915817770$\times 10^{-1}$ &\dec  2.403792997 \\
  1 & \dec 0.000000000 &\dec $-$2.149829914 &\dec  1.161082473$\times 10^{1}$ & \dec 0.000000000  &\dec $-$2.222865048 &\dec  2.332835586$\times 10^{1}$ \\
  2 &\dec $-$4.415865088$\times 10^{-1}$ &\dec $-$9.251531400$\times 10^{-1}$ &\dec  6.622899086$\times 10^{1}$ &\dec $-$4.924510549$\times 10^{-1}$ &\dec $-$1.864039733 &\dec  1.902981798$\times 10^{2}$ \\
  3 &\dec $-$2.554087280$\times 10^{-1}$ &\dec $-$6.521784931 &\dec  3.610950193$\times 10^{2}$ &\dec $-$3.192744899$\times 10^{-1}$ &\dec $-$1.517689040$\times 10^{1}$ &\dec  1.492331708$\times 10^{3}$ \\
  4 &\dec $-$5.884519404$\times 10^{-1}$ &\dec $-$8.897766802 &\dec  1.876359629$\times 10^{3}$ &\dec $-$1.446337784 &\dec $-$2.769539610$\times 10^{1}$ & \\
  5 &\dec $-$9.377312694$\times 10^{-2}$ &\dec $-$2.575353400$\times 10^{1}$ &\dec  9.557326563$\times 10^{3}$ &\dec $-$7.523886385$\times 10^{-1}$ &\dec $-$1.455619317$\times 10^{2}$ & \\
  6 &\dec $-$1.120389803 &\dec $-$4.663997073$\times 10^{1}$ & & & & \\
  7 &\dec  4.729204915$\times 10^{-1}$ &\dec $-$3.236835733$\times 10^{2}$ & & & & \\
\end{tabular}
\end{table}

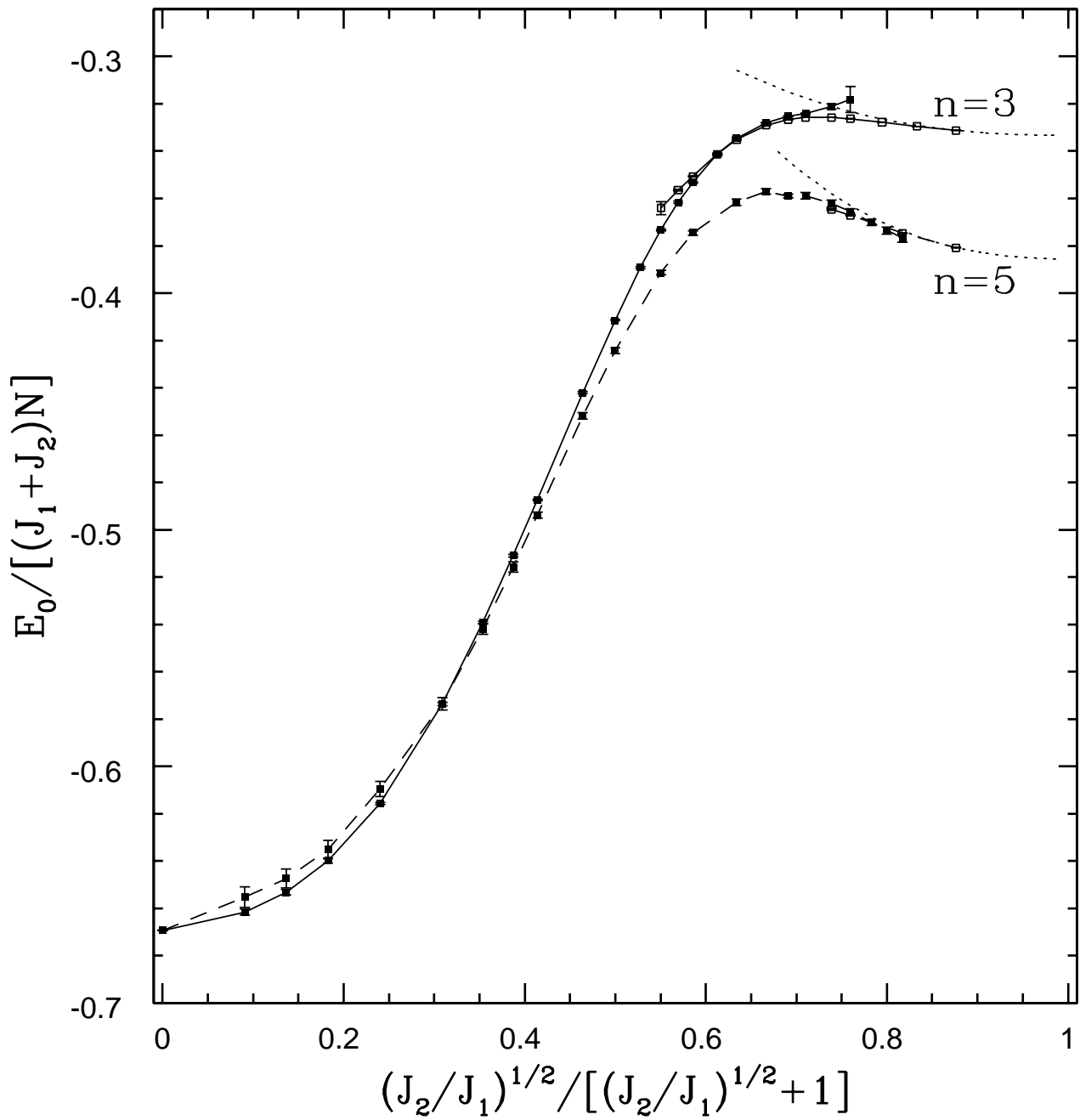
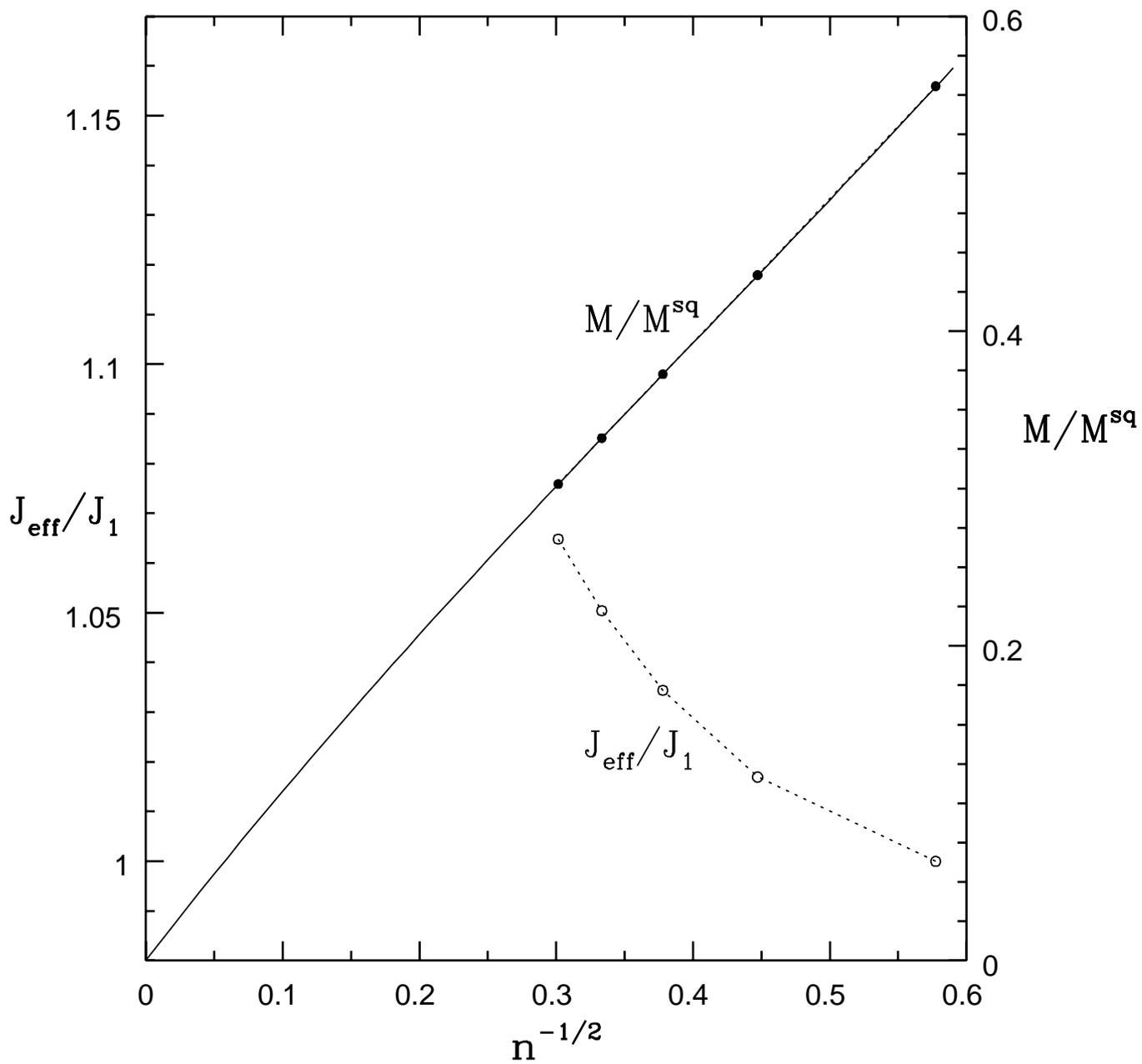
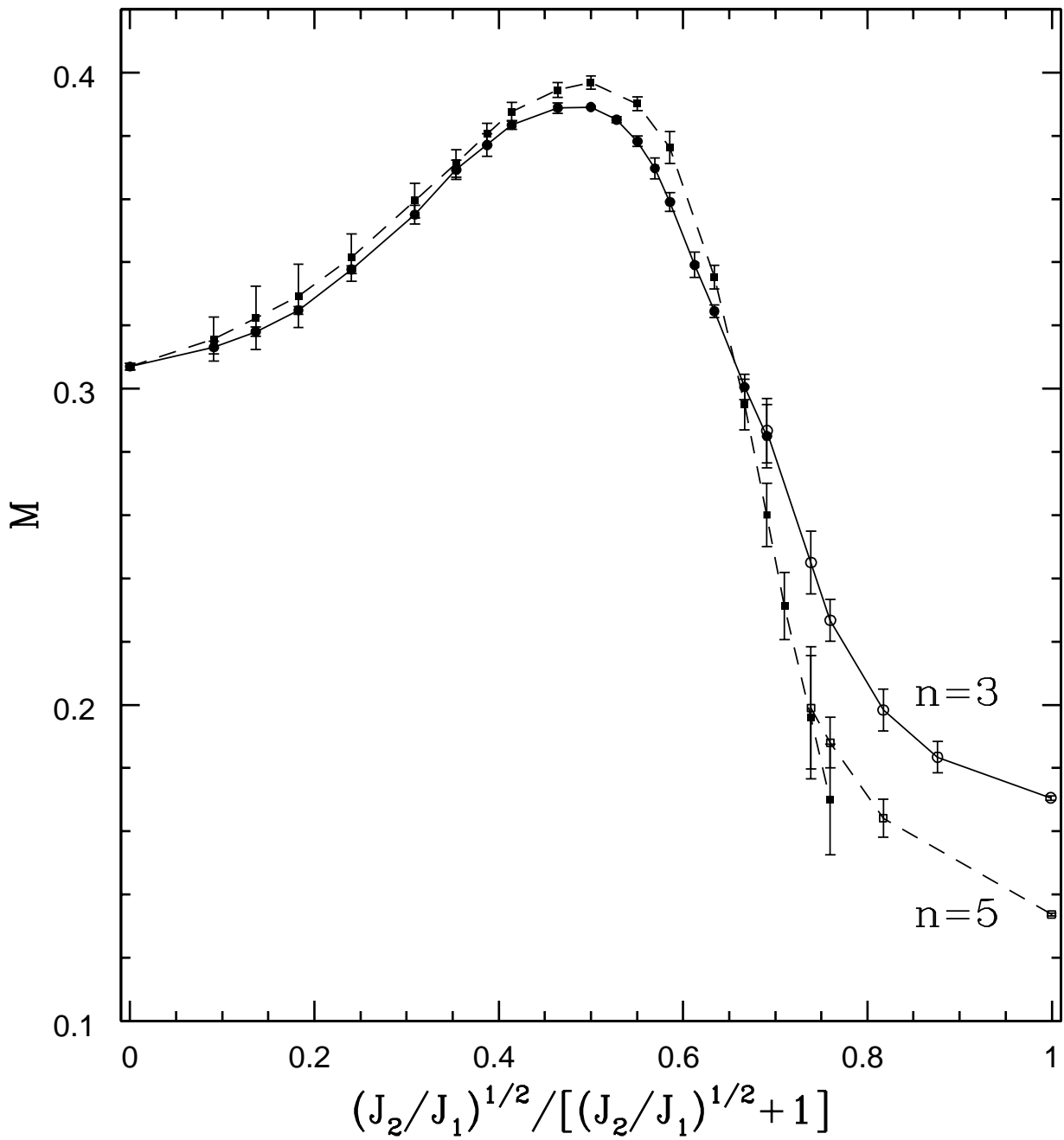
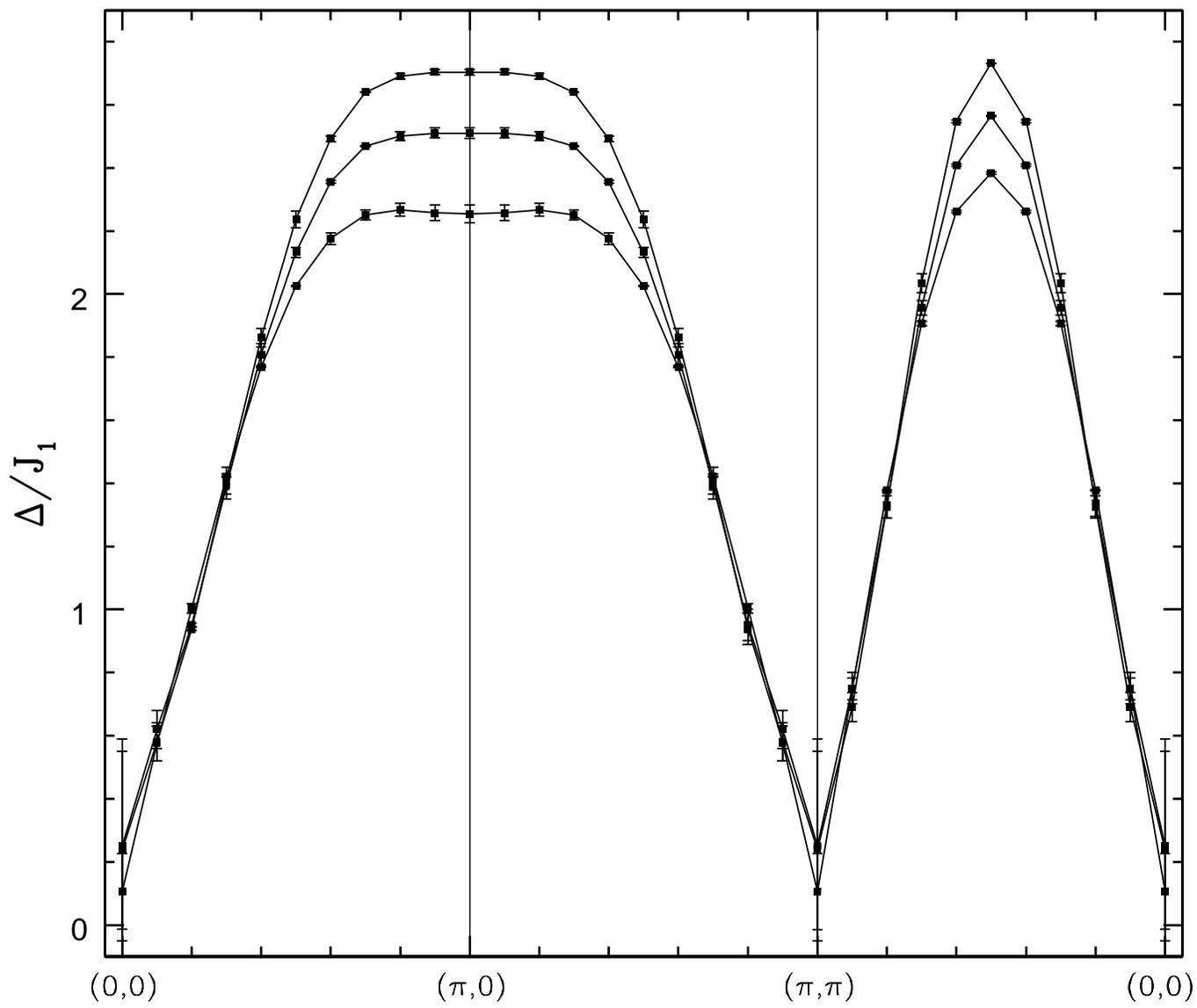
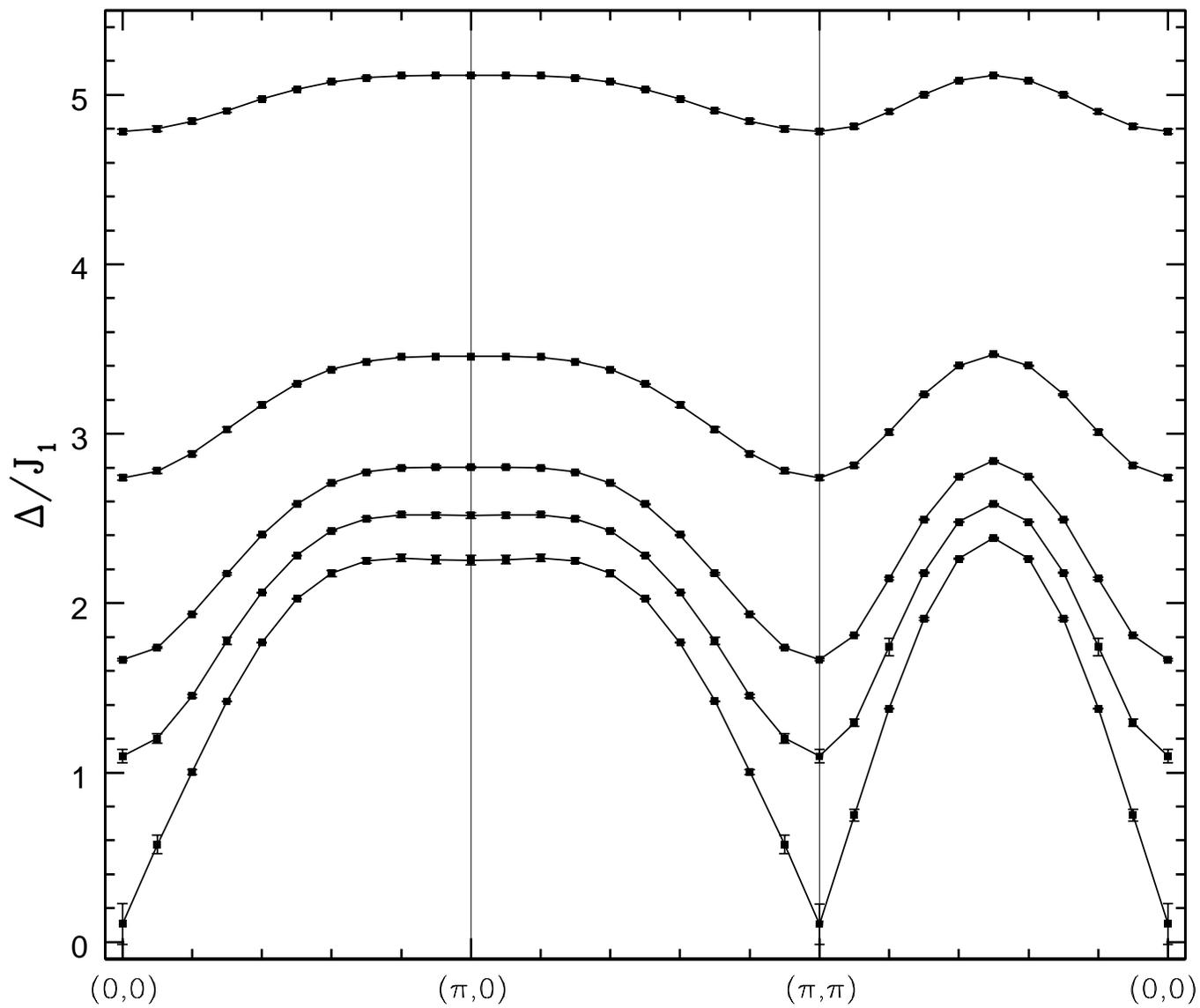
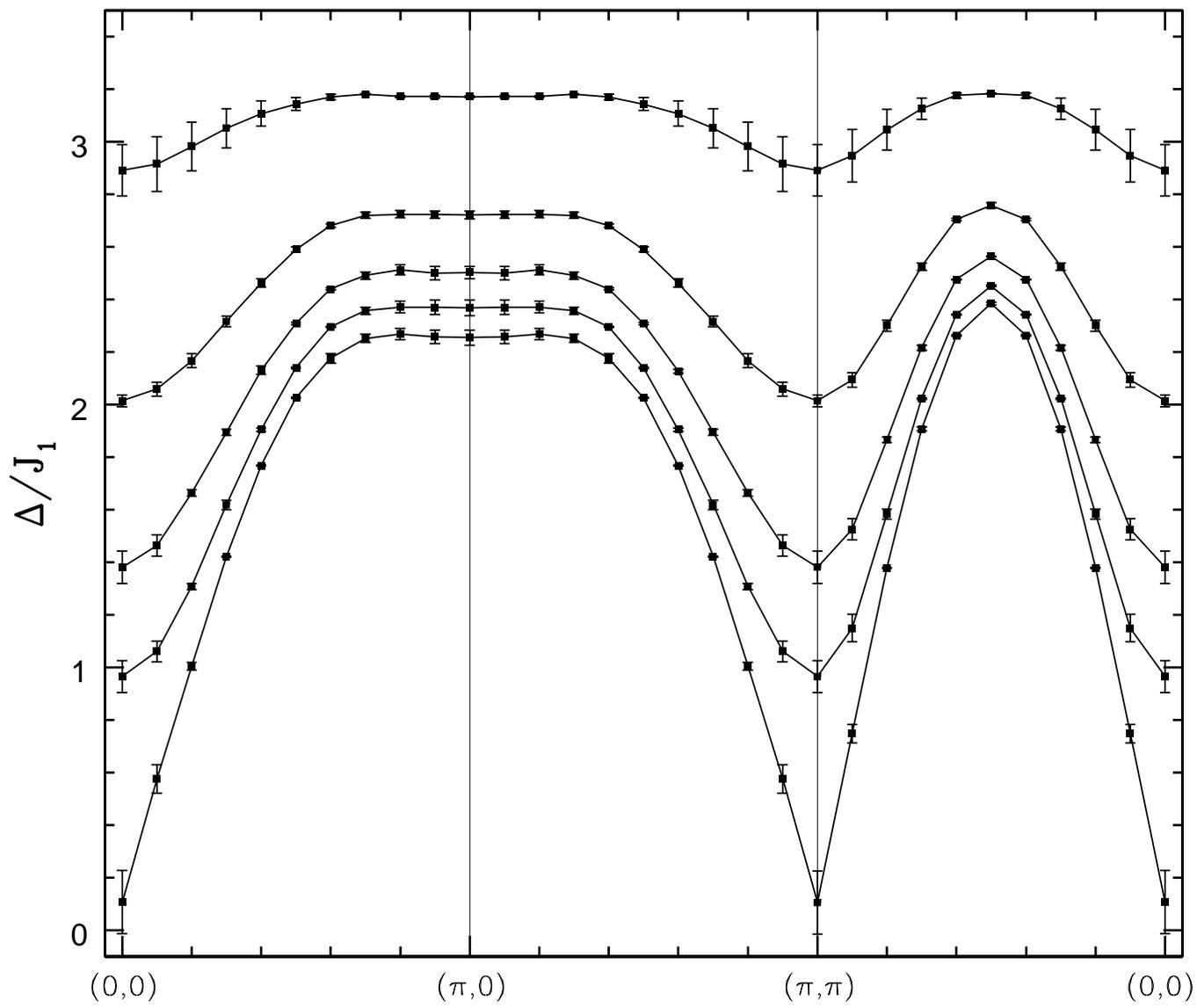
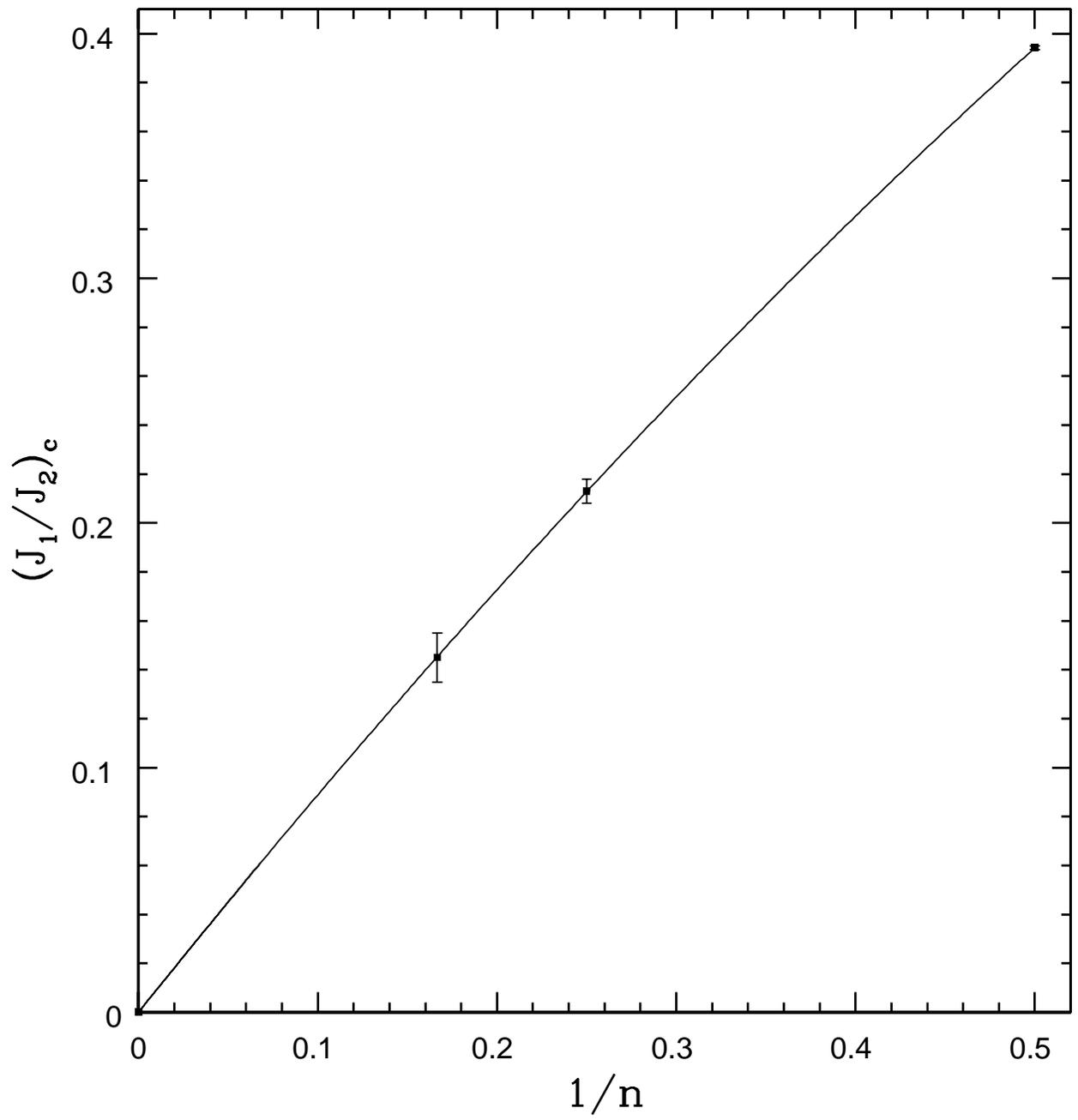
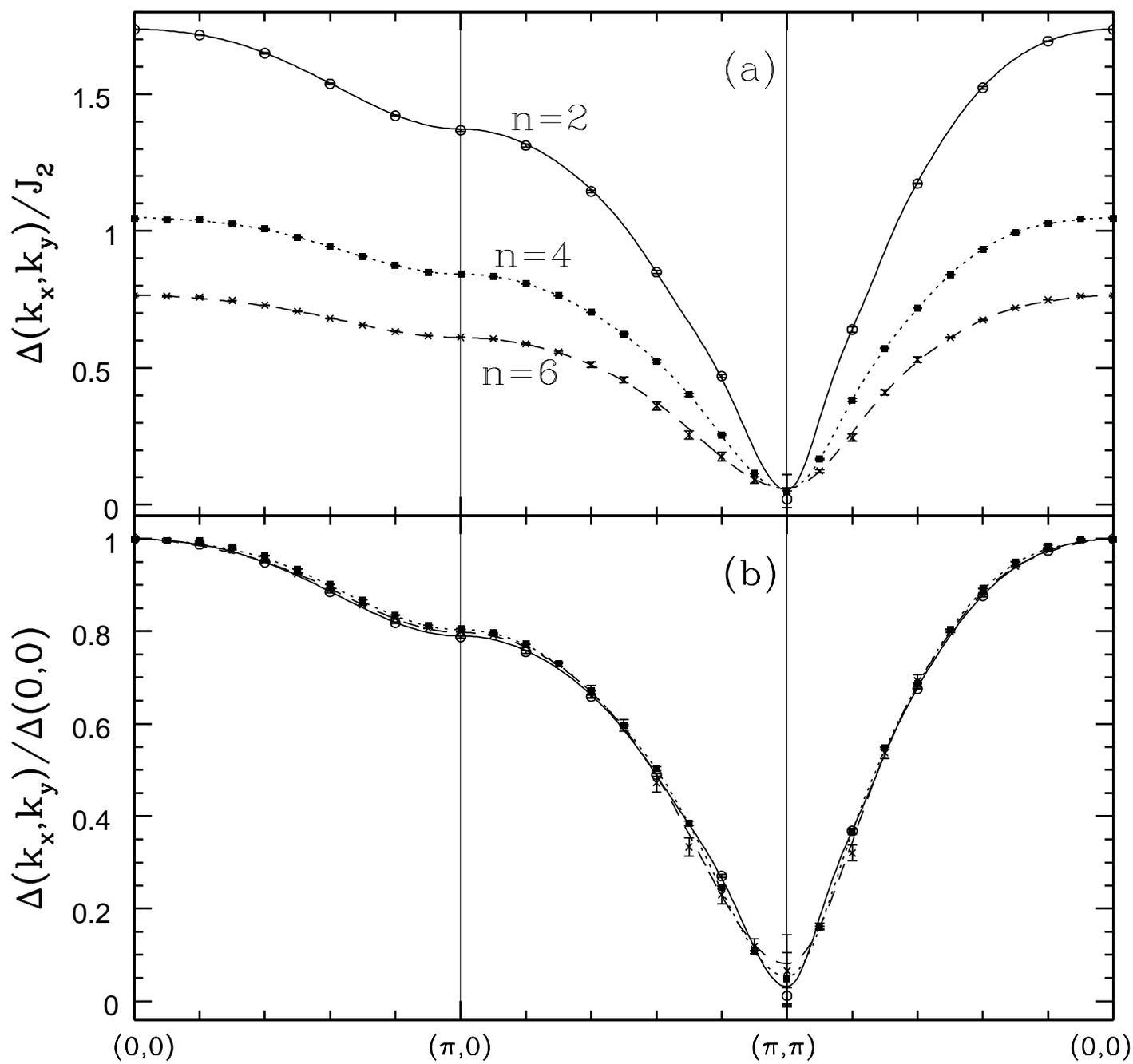
\end{document}